%% file: crowd.tex
\documentclass{sig-alternate-05-2015}
\usepackage[T1]{fontenc}
\usepackage[utf8]{inputenc}

\input{commands}
\begin{document}


\title{Early Warning of Human Crowds Based on Query Data from Baidu Map:  Analysis Based on Shanghai Stampede}
\large
\numberofauthors{1}
%
\author{Jingbo Zhou, Hongbin Pei and Haishan Wu\titlenote{Corresponding author, e-mail: wuhaishan@baidu.com.} \\
\affaddr{Baidu Research -- Big Data Lab, Beijing, China}\\
}

\maketitle
\normalsize

\begin{abstract}
Without sufficient preparation and on-site management, the mass scale unexpected huge human crowd is a serious threat to public safety. A recent impressive tragedy is the 2014 Shanghai Stampede, where 36 people were killed and 49 were injured in celebration of the New Year's Eve on December 31th 2014 in the Shanghai Bund. Due to the innately stochastic and complicated individual movement, it is not easy to predict collective gatherings, which potentially leads to crowd events. In this paper, with leveraging the big data generated on Baidu map, we propose a novel approach to early warning such potential crowd disasters, which has profound public benefits. An insightful observation is that, with the prevalence and convenience of mobile map service, users usually search on the Baidu map to plan a routine. Therefore, aggregating users' query data on Baidu map can obtain priori and indication information for estimating future human population in a specific area ahead of time.
Our careful analysis and deep investigation on the Baidu map data on various events also demonstrates a strong correlation pattern between the number of map query and the number of positioning users in an area.
Based on such observation, we propose a decision method utilizing query data on Baidu map to invoke warnings for potential crowd events about $1\thicksim3$ hours in advance.
Then we also construct a machine learning model with heterogeneous data (such as query data and mobile positioning data) to quantitatively measure the risk of the potential crowd disasters. We evaluate the effectiveness of our methods on the data of Baidu map.
\end{abstract}
\printccsdesc

\keywords{Emergency early warning; Crowd anomaly prediction; Map query; Stampede}

\input{section/intro}

\input{section/casestudy}

\input{section/prevent}

\section{Conclusion}\label{sec:conclusion}

The management of unexpected huge human crowd events is a challenging but serious problem for public safety, whereas poor management of such events may led to unfortunate deadly stampedes. With the help of the big data generated on  Baidu map, we propose a solution for improving the effectiveness of the crowd management. The novelty of the solution lies in an objective fact that many people will search on Baidu map to plan their itineraries. This information indicated by anonymous users of Baidu map provides promising and compelling advantages for us to invoke early warning for possible crowd events. Thus, we deploy a deep study for the crowd anomaly prediction with resorting to the help of the Baidu map data. Our proposed solution includes a qualitative decision method to perceive the crowd anomaly as well as a quantitative prediction method to assess the risk of the crowd event. We believe the successful deployment of our method can bring many benefits to our society.

\begin{figure}[!htb]
\centering{}
\includegraphics[width=0.45\textwidth]{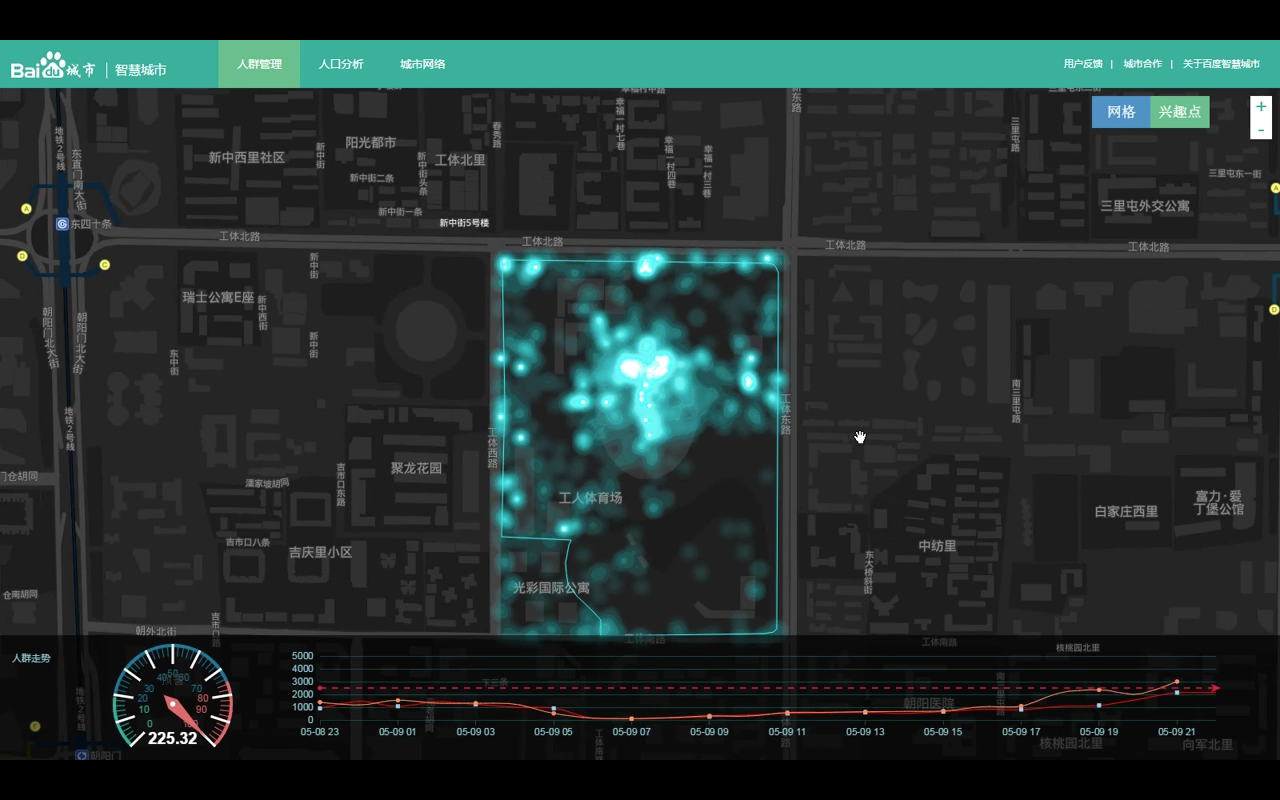}
\caption{A snapshot of the early warning system for human crowds} \label{fig:snapshot}
\end{figure}

It is worth mentioning that we have developed an early warning system for human crowds based on the aforementioned decision method and prediction model. A complete demonstration video of the system can be visited\protect\footnotemark[5], and a snapshot is shown in Figure \ref{fig:snapshot}.

\footnotetext[6]{{http://www.iqiyi.com/w\%5f19rrpwo1mt.html\#vfrm=2-3-0-1}}

\bibliographystyle{abbrv}
\bibliography{crowd}  

\end{document}

%% file: commands.tex
\usepackage{amsmath}
\usepackage{amssymb}
\usepackage{color}

\usepackage{subfigure}

\usepackage{graphicx}
\usepackage{grffile}
\usepackage{csquotes}
\usepackage{url}
\usepackage{epstopdf}
\usepackage{float}

\usepackage{multirow}

\usepackage[ruled,vlined,boxed,linesnumbered]{algorithm2e}
\usepackage{algorithmic}

\newtheorem{lemma}{ \hspace{-12pt}Lemma}[section]

\usepackage[
  pass,
]{geometry}

\usepackage{multirow}

\usepackage{hyphenat}
\usepackage[british]{babel}

\usepackage[normalem]{ulem}
\usepackage{soul}
\usepackage{xcolor}

\usepackage{float}

%% file: section/intro.tex
\section{Introduction}

The management of human crowd in public events is significantly important for public safety.
The 2014 Shanghai stampede, where 36 people were killed and 49 were injured in celebration of the
New Year's Eve on December 31th 2014, was blamed for insufficient
preparation and poor on-site management of
local officials \cite{xinghua}. Similar to other crowd disasters, one of major reasons of this tragedy is due to the wrong estimation
of the human flow and human density by management officials \cite{xinghua}\cite{pretorius2013large}.
As the potential risk is increasing rapidly with the increment of
unexpected mass scale huge crowd, inadequate emergency response
preparedness may lead to a disaster.  It is clear that the primary factor in assuring a safe
environment for crowd is a feasible emergency action plan for potential crowd events, which can help preventing such terrible crowd disasters from happening.


Preventing crowd disasters is a challenging task which
essentially depends on accurately predicting crowd anomaly.
Although many exiting works have tried to model and
predict individuals' trajectory, most of them
focus on predicting human daily routine trajectory \cite{ashok1996estimation}\cite{schneider2013unravelling}\cite{zheng2012unsupervised}\cite{zhou2013semilazy}.
However crowd anomaly is generally caused
by infrequent collective activities (e.g. celebration,
religious gatherings and sports tournaments).
Participating these activities is a non-routine but stochastic behavior
for most people, therefore, the movement of participators
is much different from their daily
routine and more like a random walk
\cite{gonzalez2008understanding}. Thus, there is no existing approach to handling such crowd anomaly prediction very well by far, to the best of our knowledge.

Traditional method for crowd anomaly detection is
based on video sensor and computer vision technology \cite{helbing2007dynamics}\cite{gallup2012visual}\cite{ali2008floor}\cite{harding2010early}.
Whereas, three limitations of this kind of methods cannot
be overcome: 1) they are too sensitive to noise of visual environment,
e.g. rapidly illumination changing may let them fail;
2) the areas with deployed video sensors are very limited in a
city; and 3) the detection methods cannot provide significant
long enough preparation time (e.g. hours level) for emergency management before crowd anomaly happening.

\begin{figure}[!htb]
\centering{}
\includegraphics[width=3.4 in]{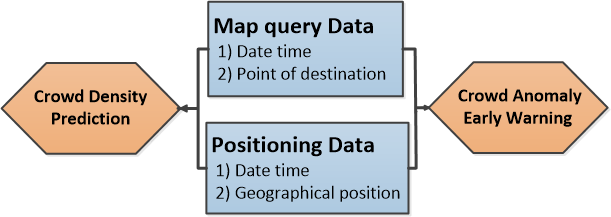}
\caption{The framework of early warning for crowd anomaly based
on map query data and positioning data.}
\label{illustration}
\end{figure}

In this work, we propose an effective early warning method for crowd anomaly,
as well as a machine learning model for crowd density prediction. Both of them are derived from
 a novel prospect by leveraging query data from Baidu map, the largest mobile map app in China provided by Baidu \footnote{http://map.baidu.com/}.
Inspired by a widespread custom that people
usually plan their trip on Baidu map before departure,
we find a strong correlation pattern between the number of
map query and the number of subsequent positioning users in
an area. That is to say, map query behavior
in some sense is a nice leading indicator and predictor for crowd dynamics.
Therefore, We divide the early warning task into two parts:
first, detecting map query dynamics of an area and sending early warning
when there is a map query number anomaly,
and then quantitatively predicting crowd density from
historical map query data and mobile positioning data of this area.
The framework is illustrated in figure \ref{illustration}.
Our approach can reliably provide early warning
for crowd anomaly for 1-3 hours
which would facilitate an early intervention for administrative agency
to prevent a crowd disaster from happening. Besides, it is worth noting that there is no user privacy issue for the proposed method since we only use the aggregated
information on the Baidu map and it is impossible to use this information to identify any individual user.

The main contributions of this work are listed as follows:
\begin{itemize}
\item We provide an inspiring perspective to predict the crowd anomaly
according to mass map query behavior.
\item We propose a simple but effective early warning method for
crowd anomaly based on map query data and positioning data.
Case studies and experiments on real data demonstrate that the proposed method
can achieve an early warning for anomaly 1-3 hours ahead.
\item We develop a machine learning model for
crowd density prediction based on historical map query data
and positioning data. Experiments on real data show that
the model can accurately predict crowd density in a specific area one hour ahead.
\end{itemize}

The rest of this paper is organized as follows.
First, we give a detailed analysis of the 2014 Shanghai Stampede
based on positioning data and map query data in Section \ref{sec:casestudy}.
Then, the details of the proposed model and the experimental results are described in
Section \ref{sec:preventing}.
Finally, we conclude this work in Section \ref{sec:conclusion}.





%% file: section/casestudy.tex
\section{Case study of the Shanghai Stampede}\label{sec:casestudy}

\subsection{Observations from the data of the Shanghai Stampede}
In this section, we present several important observations from the human mobility data about the 2014 Shanghai Stampede obtained from Baidu map, which can provide some insights for human flow and crowd anomaly prediction. The Shanghai Stampede happened at the Bund in Shanghai on Dec. 31th 2014. Hereafter, we use the ``\textbf{map query number}'' during a time interval (typically it is one hour) to denote the total number of queries whose destination is located in a specific area (like the Shanghai Bund) on Baidu map. Meanwhile, we denote the number of users {(though all the users are anonymous)} with enabling the positioning function of Baidu map during a time interval in a specific area as the ``\textbf{positioning number}''.  The position number can be treated as an approximation of the \emph{human population density} which is the number of persons per square within an area during a fixed time interval. {Note that the map query number and positioning number are only the aggregated information on Baidu map.}

\begin{figure}[!htp]
\centering{}
\includegraphics[width=0.48 \textwidth]{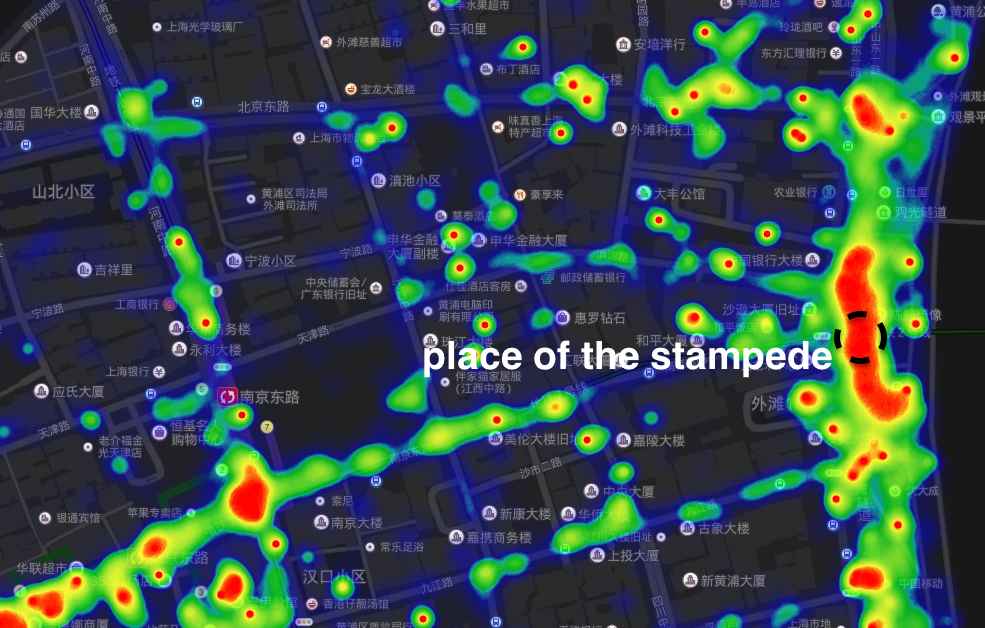}
\caption{Human population density between 23:00-24:00 on Dec. 31th 2014.}
\label{fig:heatMap_23-24}
\end{figure}

\paragraph{Observation 1: Human population density of the stampede area was higher than others}
High human population density is a necessary condition for the stampede. Figure \ref{fig:heatMap_23-24} illustrates the human density heat map between 23:00-24:00 on Dec. 31th 2014 based on the positioning number. It is obvious that the Bund, which is the disaster area of 2014 Shanghai Stampede, had higher human density than the rest parts of the city.

\begin{figure}[!htb]
\centering{}
\includegraphics[width=0.48 \textwidth]{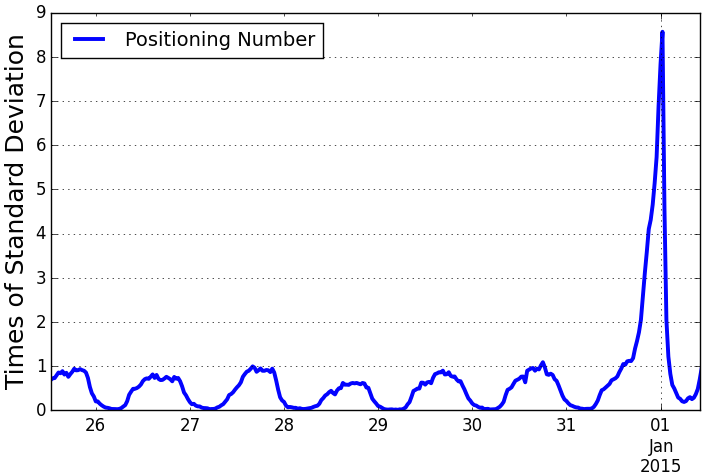}
\caption{The users' positioning number in Baidu map on the Bund  (the disaster area of 2014 Shanghai Stampede) over one week. }
\label{fig:positioning_overtime}
\end{figure}

\begin{figure}[!htb]
\centering{}
\includegraphics[width=0.48 \textwidth]{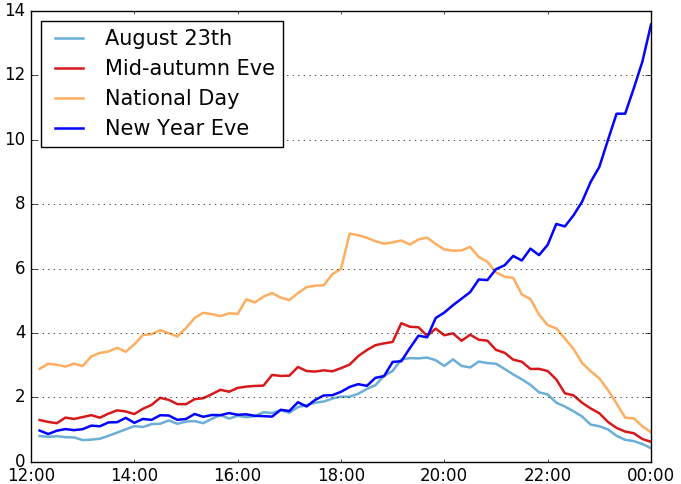}
\caption{The positioning number in Baidu map on the Bund  (the disaster area of 2014 Shanghai Stampede) in different festivals/holidays of 2014, which are a common weekend (Aug. 23th 2014), the eve of the Mid-Autumn Festival (Sept. 7th 2014, the evening before a traditional China festival), the China's National Day (Oct. 1st 2014) and New Year's Eve of 2014, respectively. All the numbers are divided by the standard deviation of the position number of the eve of the Mid-Autumn Festival.}
\label{fig:position_num_festivals}
\end{figure}

\paragraph{Observation 2: Human population density of the disaster area was higher than the ones at other times}
Human population density when the disaster happened was also higher than the ones at other times. Figure \ref{fig:positioning_overtime} shows the positioning number on Baidu map on the Bund over one week. During the disaster time, the positioning number on Baidu map was about 9 times of peak values at other times.  Furthermore, Figure \ref{fig:position_num_festivals} exhibits that the user positioning number on the Shanghai Bund of the Near Year Eve of 2014 was also higher than other festivals/holidays (In Figure \ref{fig:position_num_festivals}, the festivals/holidays are August 23th (a common weekend of 2014), the Mid-autumn eve (September 7th, the evening before a traditional China festival), the National Day (October 1st) and the New Year Eve (December 31th)).

\paragraph{Observation 3: Human flow directions of the disaster area were more chaotic}
\begin{figure*}[!htb]
\centering{}
\includegraphics[width=0.96 \textwidth]{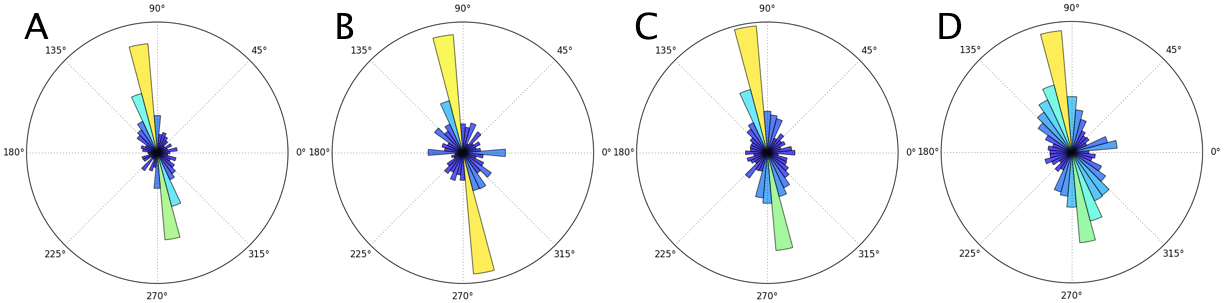}
\caption{Human flow direction distribution in Chenyi Square (the specific disaster area of 2014 Shanghai Stampede) from 22:00 to 24:00 in: A -- a common weekend (Aug. 23th 2014); B -- the eve of the Mid-Autumn Festival (Sept. 7th 2014); C -- the China's National Day (Oct. 1st 2014) and D -- New Year's Eve of 2014}
\label{fig:crowd_direction_chenyi}
\end{figure*}

We also observe that human flow directions in the disaster time were more chaotic that the ones at other times. Figure \ref{fig:crowd_direction_chenyi} illustrates the human flow directions in different festivals/holidays from 22:00 to 24:00 in the disaster area. Compared with other sub-figures, the Figure \ref{fig:crowd_direction_chenyi} (D) indicates that the disaster area had more chaos during the New Year Eve of 2014, which implies a potential accident.

\subsection{How Baidu's data can help prevent crowd disasters} \label{sec:baiduprevent}
\begin{figure}[!htb]
\centering{}
\includegraphics[width=0.48 \textwidth]{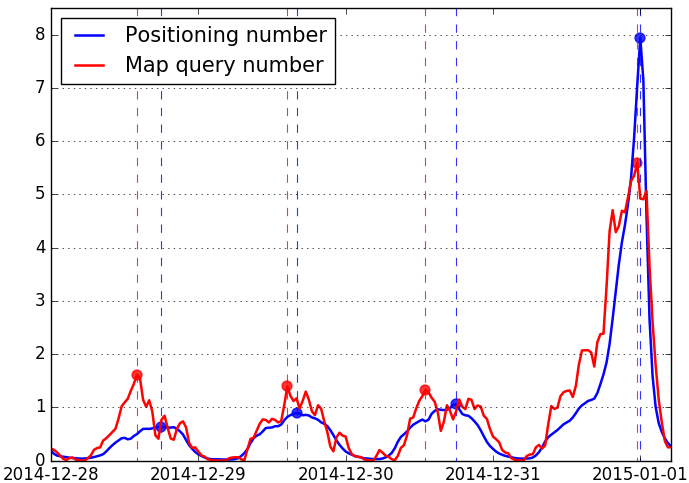}
\caption{The time series of map query number and the positioning number on Baidu map from Dec. 28th 2014 to Dec. 31th 2014 on the Shanghai Bund (Both data are normalized by dividing their standard deviation). An interesting observation is that the peak values of the map query number usually appear several hours before the peak value of the positioning number in each day.
} \label{fig:positioning_query}
\end{figure}

Our objective is to utilize the big data, especially the query data, generated by Baidu map's users to avoid such crowd disasters. If without prior information, the individual movement appears as a random walk \cite{gonzalez2008understanding}, leading to the unavoidable impediment to predict unknown human crowd event. However, nowadays before a user leaving for a destination, he or she usually uses a mobile map app to plan a routine. Therefore, aggregating and scrutinizing the massive query data on Baidu map provides an intelligent approach to predicting crowd anomaly as well as avoiding such crowd disasters. With taking more than 70\% market share overall in China, Baidu map has innate advantage to tackle this problem.

After analyzing the data on the Shanghai Stampede and other crowd events (like game events and concerts) in different places, we find that there usually are an abnormally large number of queries on Baidu map about the crowd event places emerging 0.5-2 hours before the abnormal crowd event being perceptible to human beings.

Figure \ref{fig:positioning_query} illustrates the time series of positioning number and the one of the map query number of the Baidu map in the last a few of days of 2014. This figure exposures the high correlation between the positioning number and the map query number. With a more careful inspection, we can also observe that the peak value of the map query number appears, generally, several hours before the peak value of the positioning number in each day.

\begin{figure*}[!t]
\centering
\begin{minipage}[!t]{0.45\textwidth}
\centering
\includegraphics[width=0.99\textwidth,height=0.75\textwidth]{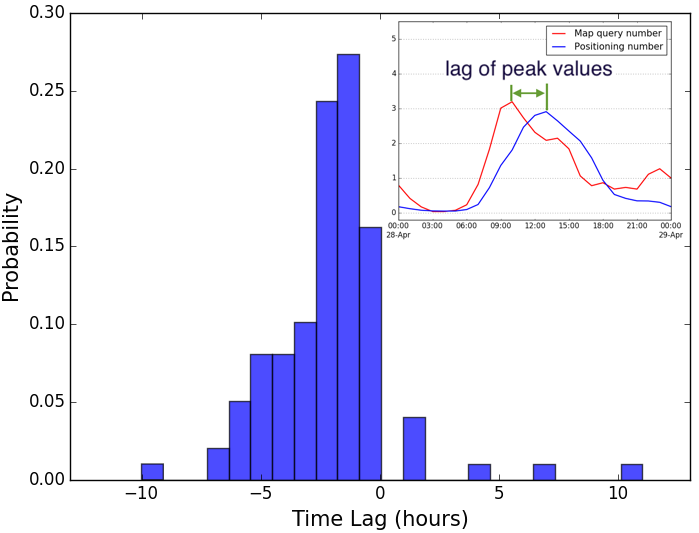}
\caption{The distribution of the lags (in hours) of the peak values between the map query number and the positioning number within the area of the Shanghai Bund in 2014. The negative lag means the peak of the map query number appears ahead of the positioning number. }
\label{fig:peaklag_waitan}
\end{minipage}
~
\begin{minipage}[!t]{0.45\textwidth}
\centering
\includegraphics[width=0.99\textwidth,height=0.75\textwidth]{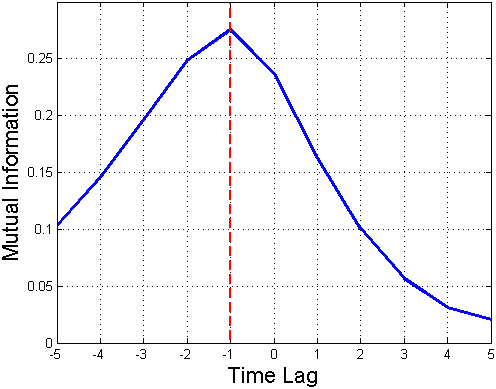}
\caption{The mutual information between the time series of the map query number and the positioning number within the area of the Shanghai Bund with different lags (in hours) in 2014. }
\label{fig:mi_waitan}
\end{minipage}
\end{figure*}

 We expound more investigations to justify the map query number as a leading indicator of human flow from the practical and and theoretical perspectives, 
 which are introduced as follows:
 \begin{itemize}

 \item First, as shown in Figure \ref{fig:peaklag_waitan}, we exhibits the distribution of the time lags (in hours) between the peak value of the map query number and the positioning number on the Shanghai Bund in each day over one year. The negative lag means the peak value of the map query number appears ahead of the one of the positioning number. As we can see from Figure \ref{fig:peaklag_waitan}, about 80\% of days in 2014 the peak value of the map query number appears one hour before the peak value of the positioning number, and with more than 50\% of days in 2014 the peak value of the map query number appears two hours before the one of the positioning number.
 \item Second, Figure \ref{fig:mi_waitan} theoretically demonstrates that the map query number can help predict the positioning number with the mutual information analysis. As shown in Figure  \ref{fig:mi_waitan}, knowing the map query number with one hour ahead can reduce the uncertainty (a.k.a the entropy) of the positioning number, which can empower our ability to predict the future human density in an area.
\end{itemize}

\subsection{The general usability of Baidu map data}
\begin{figure*}[!htb]
\centering
\begin{tabular}{ccc}
\includegraphics[width=0.32\textwidth, height =0.25\textwidth]{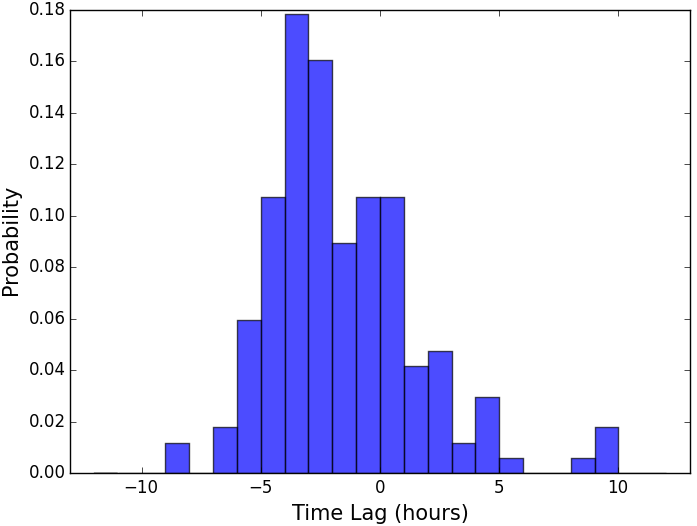} &
\includegraphics[width=0.32\textwidth, height =0.25\textwidth]{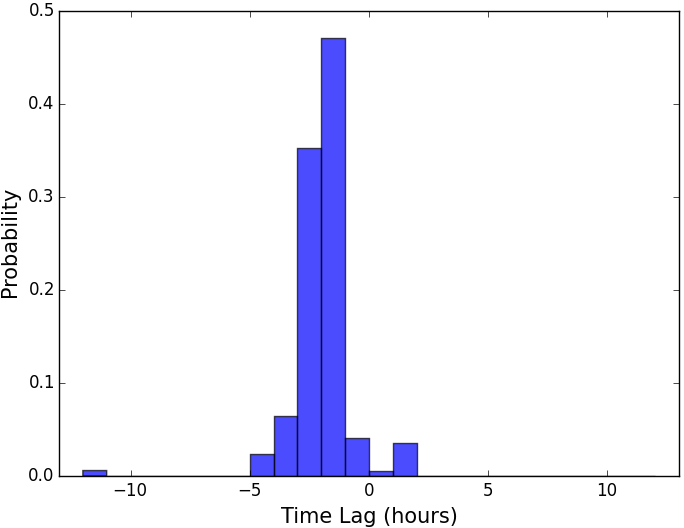} &
\includegraphics[width=0.32\textwidth, height =0.25\textwidth]{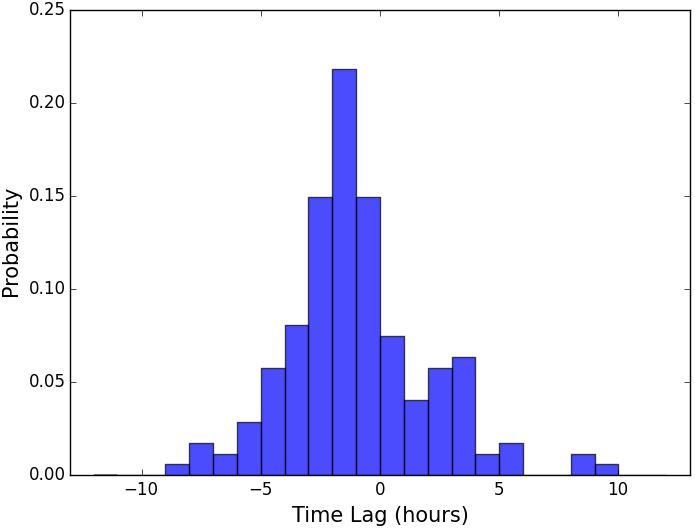} \\
\\(a) Beijing Workers' Stadium& (b) Forbidden City& 
(c) Beijing West Railway Station\\
\end{tabular}
\caption{The distribution of the lags of the peak values between the map query number and the positioning number within the area of different POIs in 2014. The negative lag means the peak of the map query number appears ahead of the one of the positioning number.}
\label{fig:peak_lag}
\vspace{5mm}
\centering
\begin{tabular}{ccc}
\includegraphics[width=0.32\textwidth, height =0.25\textwidth]{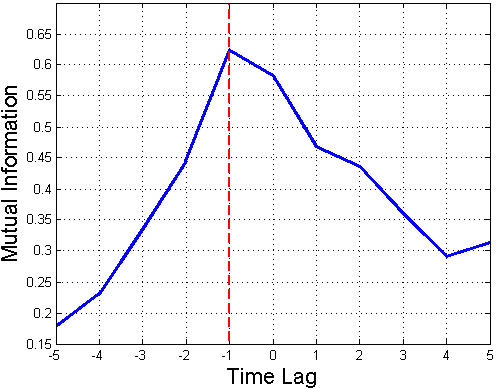} &
\includegraphics[width=0.32\textwidth, height =0.25\textwidth]{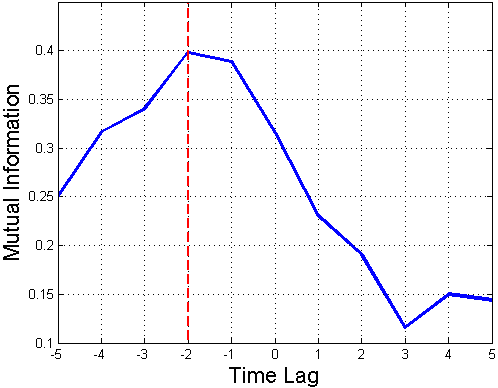} &
\includegraphics[width=0.32\textwidth, height =0.25\textwidth]{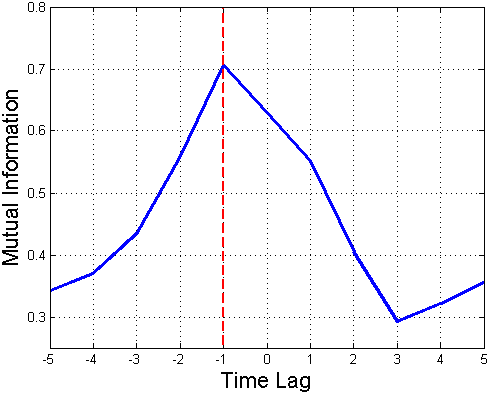} \\
\\(a) Beijing Workers' Stadium& (b) Forbidden City& 
(c) Beijing West Railway Station\\
\end{tabular}
\caption{The mutual information between the 1-year time series of the map query number and the positioning number within the area of different POIs in 2014.
}
\label{fig:mi}
\end{figure*}

We also explore the relations between the map query number and the positioning number on Baidu map in many different places to demonstrate the general usability of the map query data of Baidu map for predicting crowd anomaly. In this section, we investigate three types of such Points of Interest (POIs) in Beijing, which are: public event place, landmark, and transportation node. The selected representative places of them are  the Beijing Workers' Stadium, the Forbidden City, and the Beijing West Railway Station, respectively.

\begin{figure}[!htb]
\centering
\includegraphics[width=0.44 \textwidth]{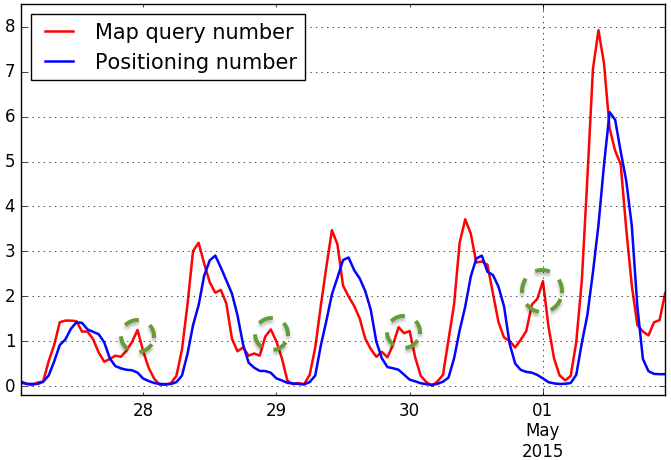}
\caption{The map query number and the positioning number of Baidu map's users from Dec. 28th 2014 to Dec. 31th 2014 in the Forbidden City (Both data are normalized by dividing their standard deviation). For each day, the time series of the map query number has two peaks before the peak of positioning number.} \label{fig:gugong_posi_query_20150501_with_circle}
\end{figure}

The statistical relation analysis, which is the same with the ones in Figure \ref{fig:peaklag_waitan} and Figure \ref{fig:mi_waitan} of Section \ref{sec:baiduprevent}, is exhibited in Figure \ref{fig:peak_lag} and Figure \ref{fig:mi}. All the analysis demonstrates that the map query number can help predict the crowd anomaly in an area. It is also worthwhile to note that different places may also have different characters for such relations. For example, in the Forbidden City, the peaks of map query number concentrate on two and three hours before the peaks of the positioning numbers. A more interesting observation of the Forbidden City is shown in Figure \ref{fig:gugong_posi_query_20150501_with_circle}. In the Forbidden city, for each day there are two peaks of the time series of the map query number before the peak of the positioning number emerging.  This reflects the complex dynamics of human behaviours and diverse environment conditions of different places.

%% file: section/prevent.tex
\section{Preventing crowd disaster with Baidu map data}\label{sec:preventing}
In this section, we detail how to utilize the Baidu map data to prevent the potential crowd disasters. We first present a qualitative decision method for crowd anomaly prediction solely based on the map query data on Baidu map. Then we construct a machine learning model with heterogeneous data to quantitatively measure the risk of the potential crowd disasters.

\subsection{A decision method for crowd anomaly prediction with map query data}

The general idea of the decision method for crowd anomaly prediction is that, if the number of map query per hour of a specific area is larger than a warning line (a certain threshold), we will invoke an anomaly warning about the possible crowd event in the following serval hours. This decision method is based on relation analysis between the map query number and the positioning number discussed in Section \ref{sec:casestudy}, which indicates that a large number of map query number subsequently imply a potential large number of human population in a few of hours in a specific area .

The warning line for the map query number is statistically computed with the peak values of the map query number per hours in every days under the log-normal probability distribution model. Formally, let us denote the time series of map query number per hours on the Baidu map by $M={m(t)}, t\in \mathbb{N}_0$. For a given date $d$, we first calculate the peak value of the map query number as $pm(d)$:
\begin{equation*}
pm(d) = \max_{t=24d}^{24(d+1)}m(t)
\end{equation*}
In our method, we consider $pm(d)$ as random variable and assume it follows log-normal distribution, i.e.:
\begin{equation}
\varphi(d) = log(pm(d)) \sim N(\mu_{pm},\sigma_{pm}^2)
\end{equation}
We use one year time series data of map query number to obtain an unbiased estimation of $\mu_{pm}$ and $\sigma_{pm}$ which are sample mean $\hat{\mu}_{pm}$ and sample variance $\hat{\sigma}^2_{pm}$:
\begin{align}
\hat{\mu}_{pm} &= \frac{1}{365}\sum_{d=0}^{364}log(pm(d))\\
\hat{\sigma}^2_{pm} &= \frac{1}{365-1}\sum_{d=0}^{364} (log(pm(d))-\hat{\mu}_{pm})^2
\end{align}
Accordingly, the warning line of the map query number is determined as
\begin{equation}\label{fig:omega_m}
\omega_m=\hat{\mu}_{pm} + \alpha \hat{\sigma}_{pm},
\end{equation}
where $\alpha$ is a model parameter and $\alpha>0$. The effect of $\alpha$ is evaluated in Section \ref{sec:exp_deccision}.

Similarly, we can also obtain a warning line of the positioning number. We define the time series of position number as $Q=\{q(t)\}$, and the peak value of positioning number for date $d$ is $pq(d)$:
\begin{equation*}
pq(d) = \max_{t=24d}^{24(d+1)}q(t)
\end{equation*}
We also assume that $pq(d)$ follows log-normal distribution, i.e.:
\begin{equation}
\phi(d) = log(pq(d)) \sim N(\mu_{pq},\sigma_{pq}^2)
\end{equation}
and we have the sample estimation of $\hat{\mu}_{pq}$ and $\hat{\sigma}^2_{pq}$:
\begin{align}
\hat{\mu}_{pq} &= \frac{1}{365}\sum_{d=0}^{364}log(pq(d))\\
\hat{\sigma}^2_{pq} &= \frac{1}{365-1}\sum_{d=0}^{364} (log(pq(d))-\hat{\mu}_{pq})^2
\end{align}
In this paper, the warning line of positioning number is fixed as $\omega_q=\hat{\mu}_{pq}+3\hat{\sigma}_{pq}$.

Finally, our decision method can be formally described by following lemma:
\begin{lemma}\label{lemma:decision}
If $m(t)\geq \omega_m$, then we have $q(t+\Delta) \geq \omega_q$ and $1\leq \Delta\leq T$ with high probability, where $T$ is a limited time period.
\end{lemma}
In our following demonstrations and evaluations, we set $T=3$ (hours).

\subsubsection{Demonstration of the decision method}


\begin{figure*}[!pb]
\small
\centering
\begin{tabular}{ccc}
\includegraphics[width=0.32\textwidth, height =0.19\textwidth]{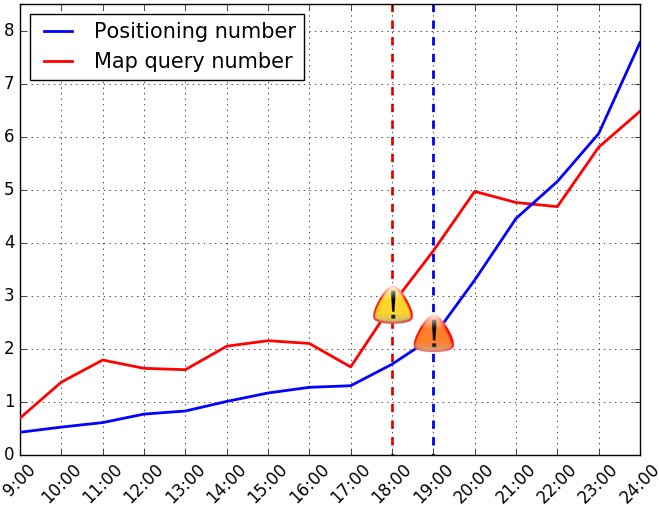} &
\includegraphics[width=0.32\textwidth, height =0.19\textwidth]{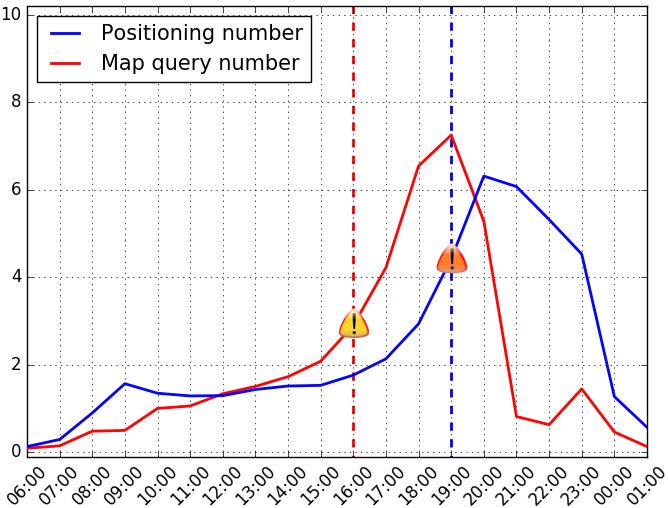} &
\includegraphics[width=0.32\textwidth, height =0.19\textwidth]{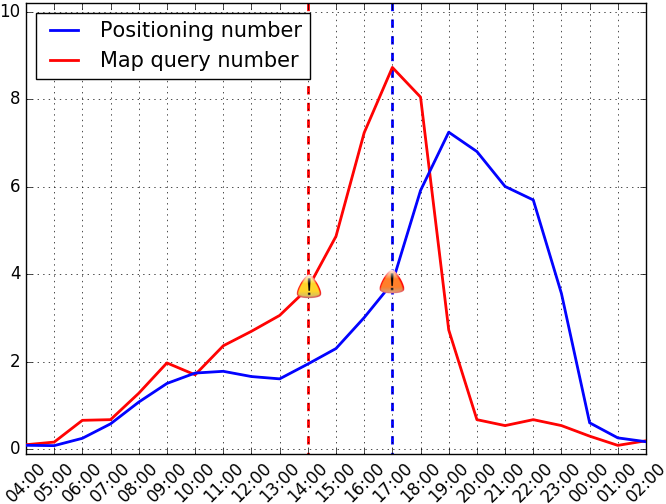} \\
(a) Shanghai Bund in 2014/12/31& (b) Shanghai Stadium in 2015/07/30&
(c) Shenzhen Bay Sports Center in 2015/04/25\\
\end{tabular}
\caption{An illustration of the decision method for crowd anomaly prediction with map query data on Baidu map in Shanghai and Shenzhen.  If the map query number is larger than a warning line, we will invoke a warning for crowd anomaly in the following several hours (see Lemma \ref{lemma:decision}). From the figure we can see that the event warning can be sent 1$\sim$3 hours before the positioning number surpassing its warning line. The corresponding events are (a) 2014 Shanghai Stampede, (b) the International Champions--Real Madrid vs AC Milan\protect\footnotemark[1] and (c) TVXQ! special live tour T1story in Shenzhen \protect\footnotemark[2]. }
\label{fig:decision_shanghai}
\vspace{5mm}
\centering
\begin{tabular}{ccc}
\includegraphics[width=0.32\textwidth, height =0.19\textwidth]{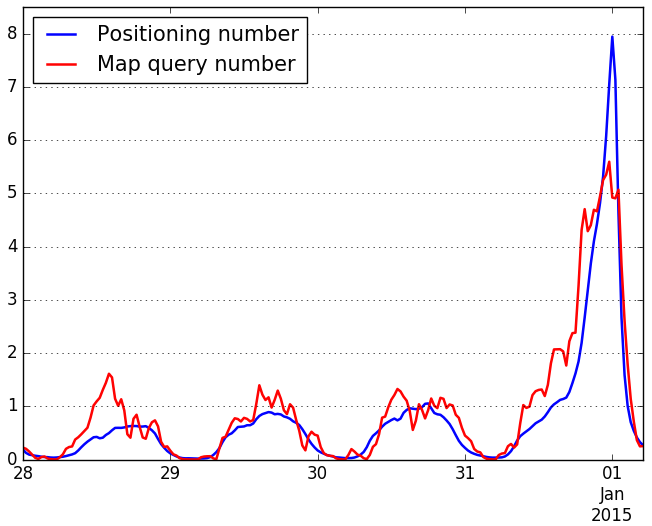} &
\includegraphics[width=0.32\textwidth, height =0.19\textwidth]{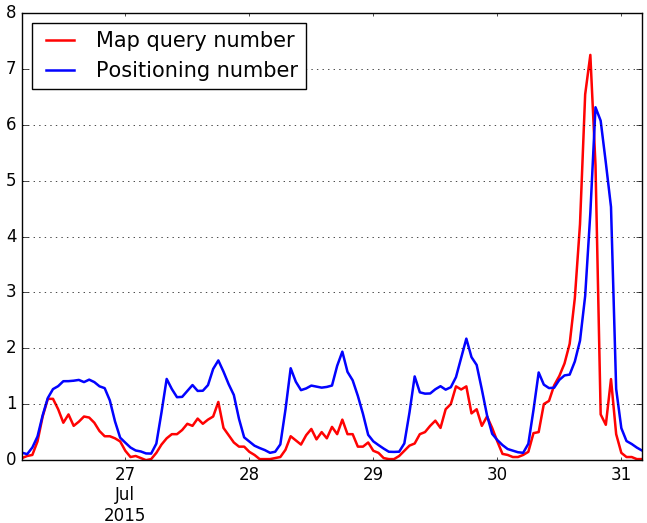} &
\includegraphics[width=0.32\textwidth, height =0.19\textwidth]{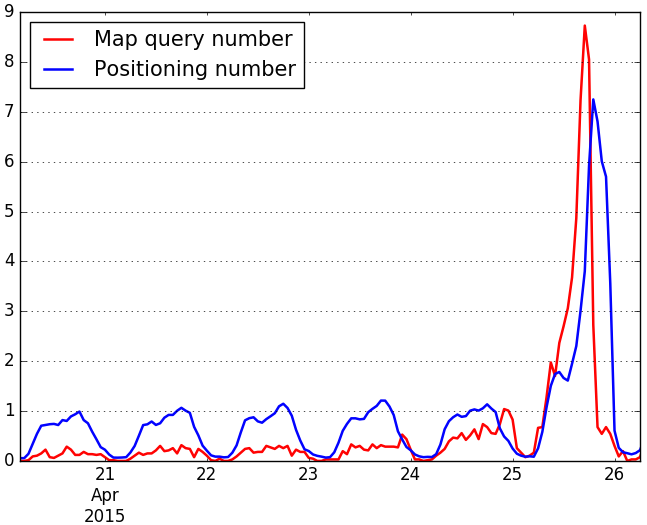} \\
(a) Shanghai Bund in 2014/12/31& (b) Shanghai Stadium in 2015/07/30&
(c) Shenzhen Bay Sports Center in 2015/04/25\\
\end{tabular}
\caption{The map query number and the positioning number on the Baidu map corresponding to the events of Shanghai and Shenzhen shown in Figure \ref{fig:decision_shanghai}.
}
\label{fig:mappos_shanghai}
\normalsize
\vspace{3mm}
\small
\centering
\begin{tabular}{ccc}
\includegraphics[width=0.32\textwidth, height =0.19\textwidth]{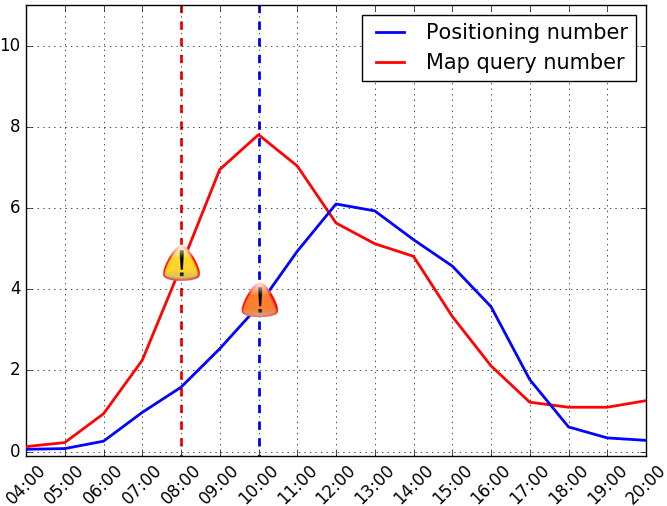} &
\includegraphics[width=0.32\textwidth, height =0.19\textwidth]{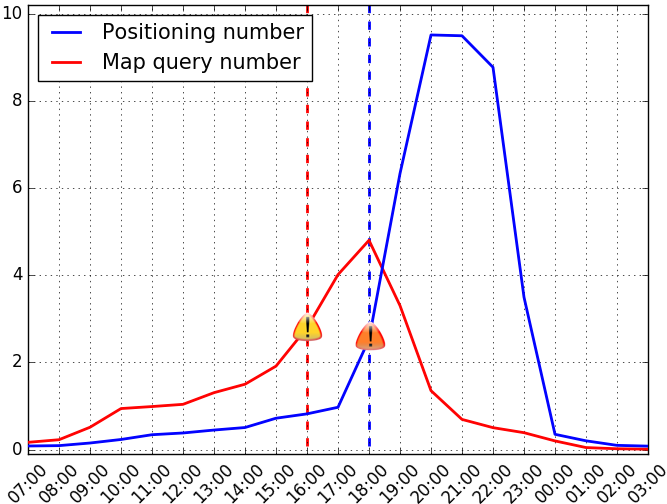} &
\includegraphics[width=0.32\textwidth, height =0.19\textwidth]{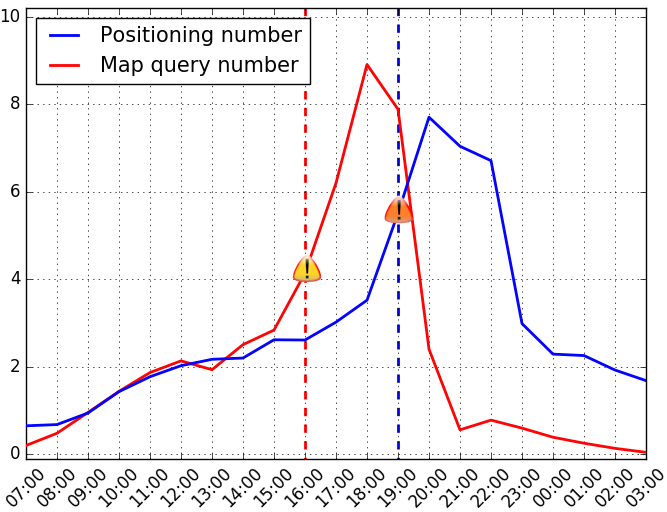} \\
(a) Forbidden City in 2015/05/01 & (b) Beijing National Stadium in 2015/10/07&
(c) Beijing Workers' Stadium in 2015/07/26\\
\end{tabular}
\caption{An illustration of the decision method for crowd anomaly prediction with map query data on Baidu map in Beijing (similar with Figure \ref{fig:decision_shanghai}).  If the map query number is larger than a warning line, we will invoke a warning for crowd event in the following several hours (see Lemma \ref{lemma:decision}). From the figure we can see that the event warning can be sent 1$\sim$3 hours before the positioning number surpassing its warning line. The corresponding events are (a) International Workers' Day, (b) Finals of the Voice of China\protect\footnotemark[3] and (c) the Hunting Party Chinese Tour\protect\footnotemark[4]. }
\label{fig:decision_beijing}
\vspace{5mm}
\centering
\begin{tabular}{ccc}
\includegraphics[width=0.32\textwidth, height =0.19\textwidth]{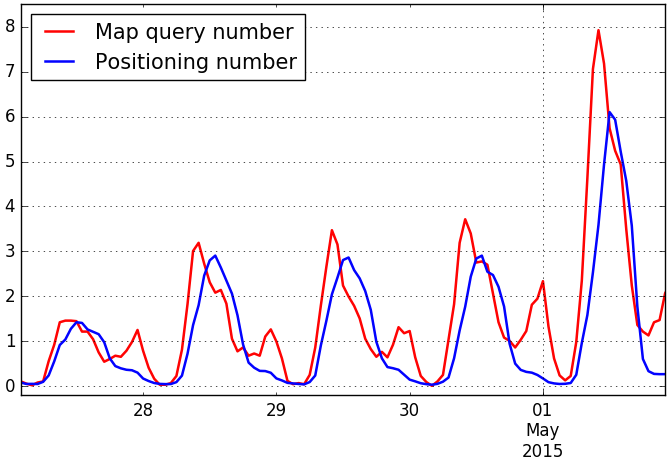} &
\includegraphics[width=0.32\textwidth, height =0.19\textwidth]{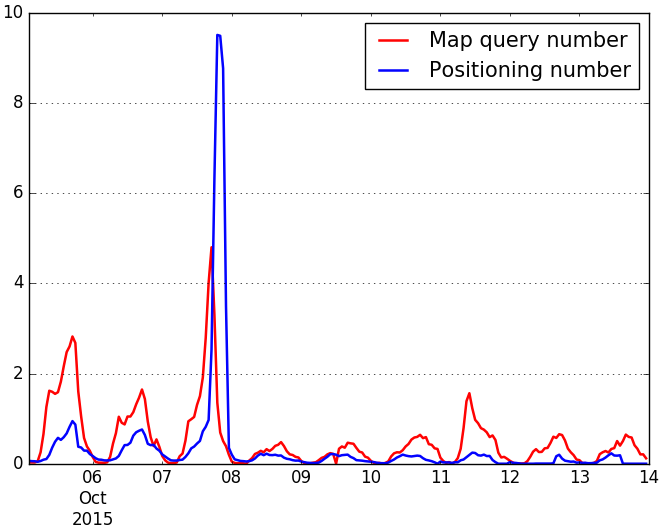} &
\includegraphics[width=0.32\textwidth, height
=0.19\textwidth]{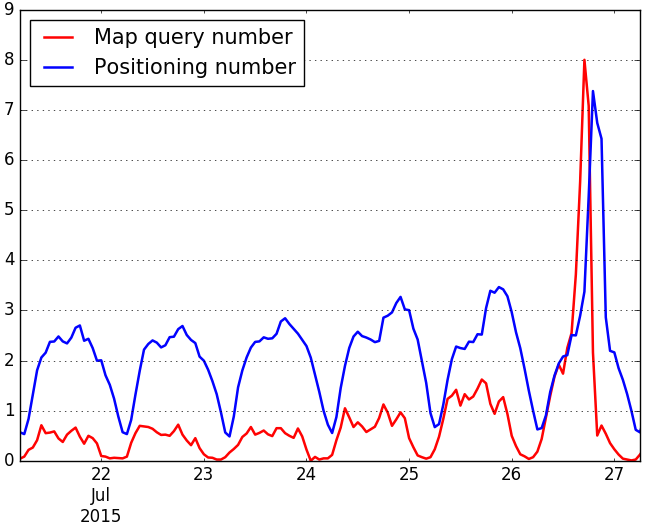} \\
(a) Forbidden City in 2015/05/01& (b) Beijing National Stadium in 2015/10/07&
(c) Beijing Workers' Stadium in 2015/07/26\\
\end{tabular}
\caption{The map query number and positioning number on the Baidu map corresponding to the events of Beijing shown in Figure \ref{fig:decision_beijing}.
}
\label{fig:mappos_beijing}
\normalsize
\end{figure*}
\footnotetext[2]{\url{http://www.goal.com/en-za/match/real-madrid-vs-milan/2032786/report}}
\footnotetext[3]{\url{http://www.springcocoon.com/html/CN/ssyc/hdgl/201504/07847.html}}
\footnotetext[4]{\url{http://www.n-s.cn/ssyc/1/}}
\footnotetext[5]{\url{http://lplive.net/shows/db/2015/20150726}}

We demonstrate several examples for crowd event prediction based on Lemma \ref{lemma:decision} in different POIs of several China cities.  Figure \ref{fig:decision_shanghai} illustrates such decision method on the data about the public crowd anomaly prediction of Shanghai and Shenzhen (which can be compared with Figure \ref{fig:mappos_shanghai}). As we can see from Figure \ref{fig:decision_shanghai}, a warning about the abnormal crowd event can be invoked $1-3$ hours before the positioning number surpassing a warning line. Figure \ref{fig:decision_beijing} demonstrates similar results for the abnormal crowd event prediction in Beijing.  To sum up, with monitoring the map query number for each hour, it is possible to give a very early warning for the abnormal crowd event to avoid crowd disasters.


\subsubsection{Performance evaluation of the decision method}\label{sec:exp_deccision}
\begin{figure*}[!htb]
\centering
\begin{tabular}{ccc}
\includegraphics[width=0.31\textwidth]{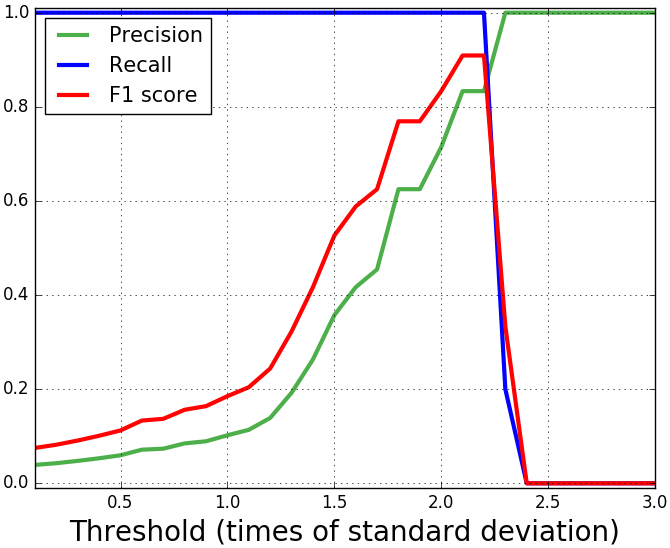} &
\includegraphics[width=0.31\textwidth]{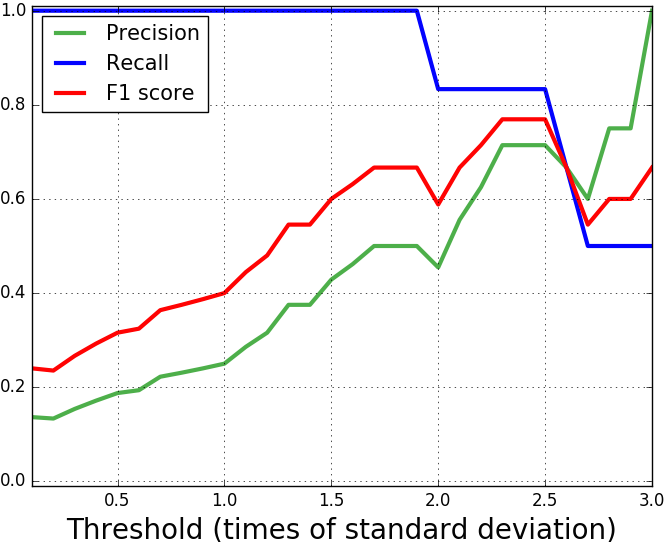} &
\includegraphics[width=0.31\textwidth]{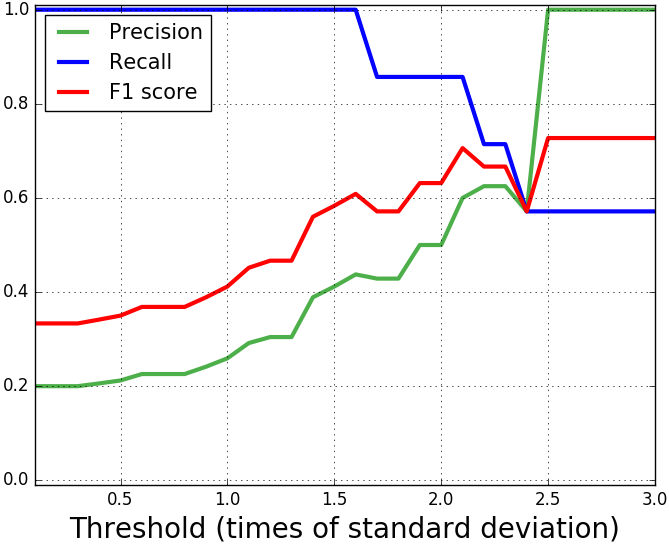}
\\(a) Shanghai Bund& (b) Shanghai Stadium&
(c) Shenzhen Bay Sports Center\\\\
\includegraphics[width=0.31\textwidth]{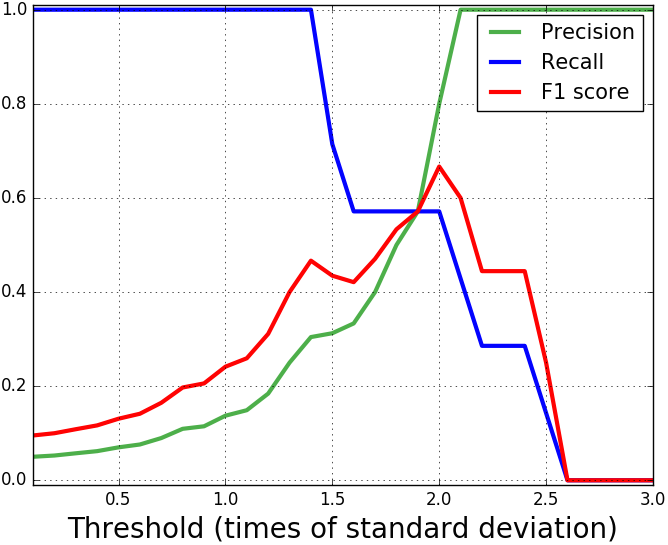} &
\includegraphics[width=0.31\textwidth]{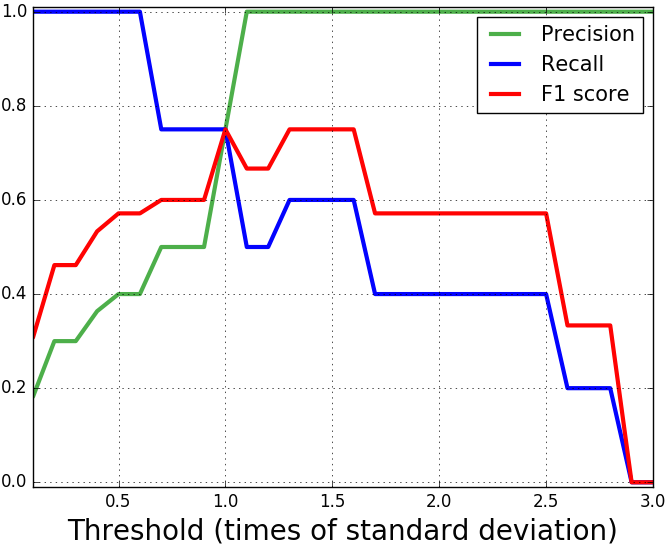} &
\includegraphics[width=0.31\textwidth]{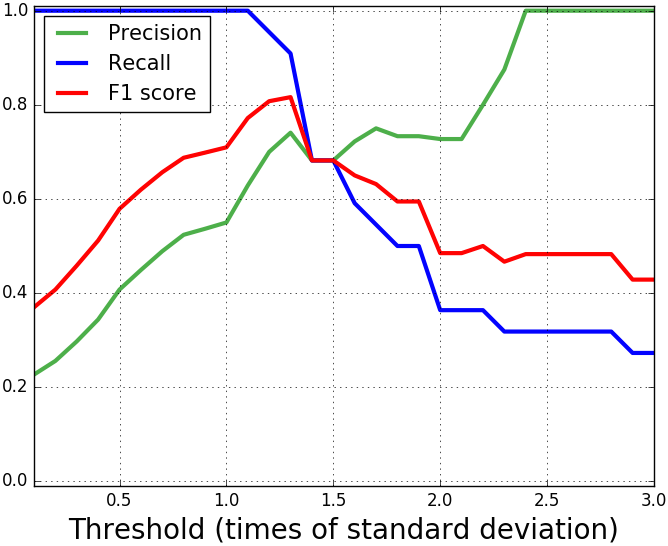} \\
(d) Forbidden City & (e) Beijing National Stadium &
(f) Beijing Workers' Stadium \\
\end{tabular}
\caption{The precision, recall and F1-score for crowd anomaly prediction with map query data in different places with different times (i.e. $\alpha$) of standard deviations.}
\label{fig:pre_recall_pois}
\end{figure*}

\begin{figure}[!htb]
\centering{}
\includegraphics[width=0.45 \textwidth]{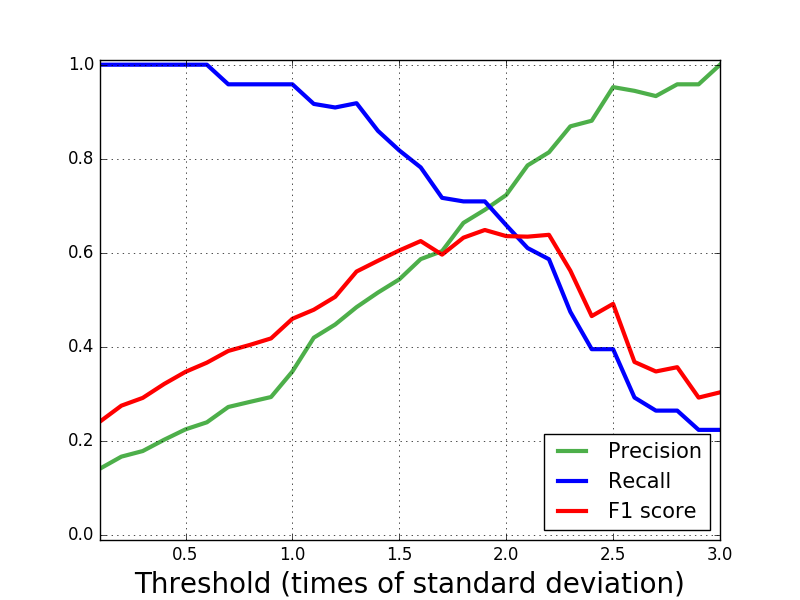}
\caption{The precision, recall and F1-score  for crowd anomaly prediction with map query data averaged in six POIs (which are Shanghai Bund, Shanghai Stadium, Shenzhen Bay Sports Center, Forbidden City, Beijing National Stadium and Beijing Workers' Stadium).} \label{fig:pre_recall_all}
\end{figure}

We also employ a performance evaluation for the decision method. In the six POIs of Figure \ref{fig:decision_shanghai} and Figure \ref{fig:decision_beijing}, we first select all the time points where $q(t)>\omega_q$, and consider these points as the ground truth of crowd events. Then we use our decision method introduced in Lemma \ref{lemma:decision} to try to predict the possible crowd events. In other words, for a monitored POI, if the map query number is larger than a predefined warning line, we will invoke a warning that there will be a crowd event in the following $1-3$ hours. We exhibit the precision, recall and F1-score of six POIs in Figure \ref{fig:pre_recall_pois}. As we can see from the Figure \ref{fig:pre_recall_pois}, there is a trade-off between the precision and recall, which is controlled by the times of the standard deviations (i.e. $\alpha$ in Eqn. (\ref{fig:omega_m})). We also aggregate the precision, recall and F1-score of the six POIs together, and illustrate the average measures in Figure \ref{fig:pre_recall_all}.


\subsection{Machine learning model for risk control of the potential crowd disasters}
We also develop a machine learning model to utilize the map query data, historical positioning data and other information available in Baidu map to predict the future human density in an area. Such predicted density can be taken as a quantitative measure to assess the potential risk of a crowd disaster. The intuition is that the probability of a crowd disasters is directly related with the human density: the human density is a necessary (though not sufficient) condition for a crowd disaster.

\subsubsection{Problem definition}

 Let us denote $M_{t_c}={m(t)}$ and $Q_{t_c}={q(t)}, t\in [0,...,t_c]$ as time series of map query number and time series of positioning number before time $t_c$ in a specific area, respectively.
We set time granularity to one hour with a fine resolution and statistical significance. And let $ f(M_{t_c},Q_{t_c})$ denote the prediction model with two time series as inputs, then the positioning number at next time $q(t_c+1)$ can be estimated as:
\begin{equation}
q(t_c+1) \approx \hat{q}(t_c+1) = f(M_{t_c},Q_{t_c})
\end{equation}
Thus, our objective is to train a prediction model $f(\cdot)$, based on historical map query data and positioning data, to accurately predict the positioning number in next time $q(t_c+1)$.

\subsubsection{Model and Feature selection}
The positioning number prediction task can be modeled as a regression problem, where the positioning number in next hour $q(t_c+1)$ is the target value.
A GBDT (Gradient Boosting Decision Tree) model,  which is an effective and widely adopted ensemble machine learning technique, is employed for the task.
In the GBDT model, we extract 47 features from two time series $M_{t_c}$ and $Q_{t_c}$ for the prediction task. The details of the features is describe in Table \ref{tab:feature}.

\small
\begin{table}[!htb]
\caption{Feature Instruction}\label{tab:feature}
\begin{tabular}[t]{l|l|l}
\hline
ID &Name & Feature describe \\
\hline
1&PN1 & positioning number 1 hour ago \\
....&... & ... \\
4& PN4 & positioning number 4 hours ago \\
5& PNS1 & positioning number at same hour 1 day ago \\
...&... & ... \\
11&PNS7 & positioning number at same hour 7 day ago \\
12 & MQ1 & map query number 1 hour ago \\
13 & MQ2 & map query number 2 hour ago \\
14 & MQY & map query number in 20:00 - 24:00  yesterday\\
15 & CT1 & current time is 1 o'clock (bool)\\
...& ... & ...\\
38 & CT24 & current time is 24 o'clock (bool value)\\
39 & TW1 & today is Monday (bool value)\\
... & ... & ...\\
45&TW7 & today is Sunday (bool value)\\
46&WD & today is weekend (bool value)\\
47&HD & today is holiday (bool value)\\
\hline
\end{tabular}
\end{table}

\normalsize

\subsubsection{Experiments}
\begin{figure*}[!htb]
\small
\centering
\begin{tabular}{ccc}
\includegraphics[width=0.32\textwidth]{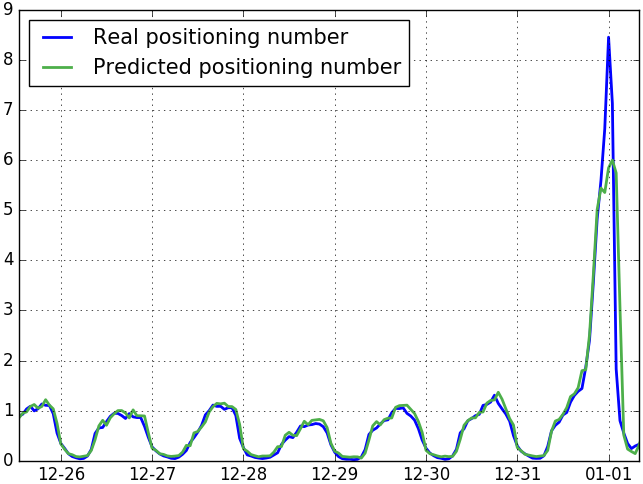} &
\includegraphics[width=0.325\textwidth]{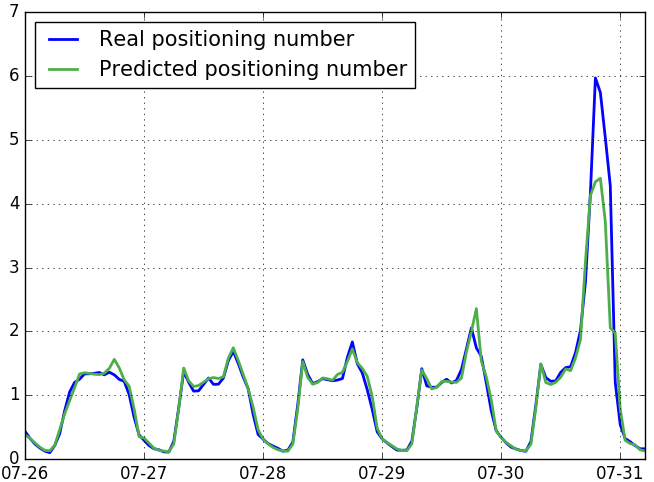} &
\includegraphics[width=0.32\textwidth]{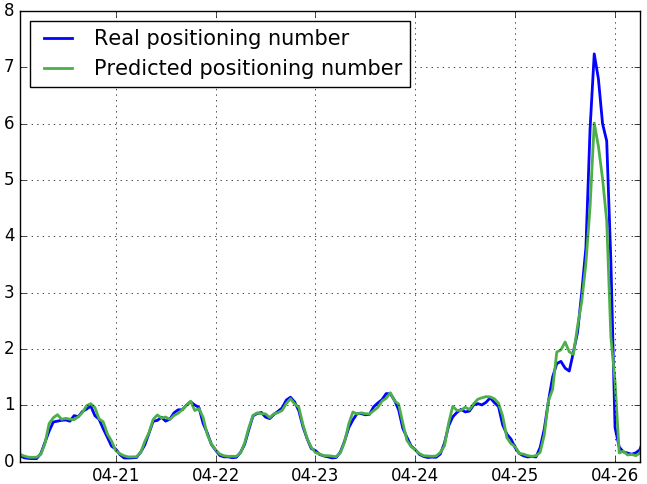}
\\(a) Shanghai Bund in 2014/12/31& (b) Shanghai Stadium in 2015/07/30&
(c) Shenzhen Bay Sports Center in 2015/04/25\\\\
\includegraphics[width=0.325\textwidth]{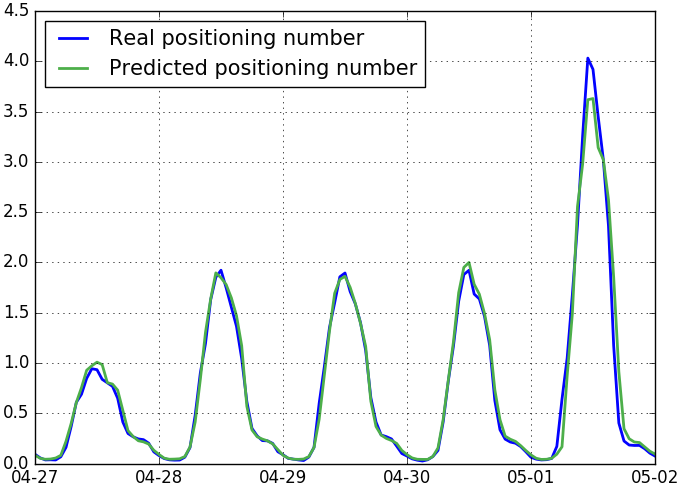} &
\includegraphics[width=0.325\textwidth]{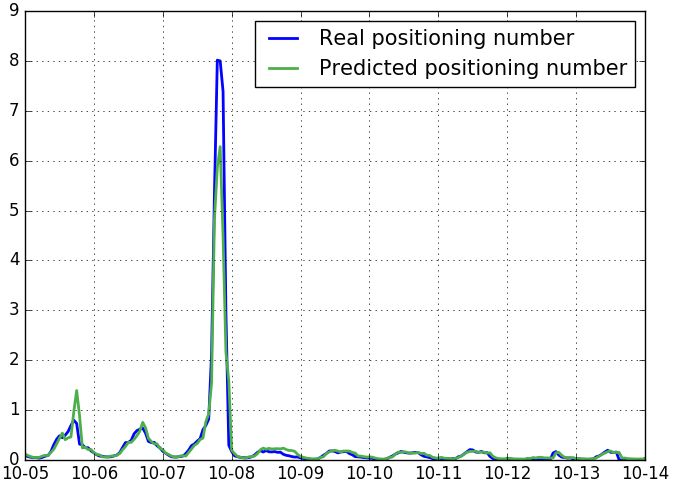} &
\includegraphics[width=0.32\textwidth]{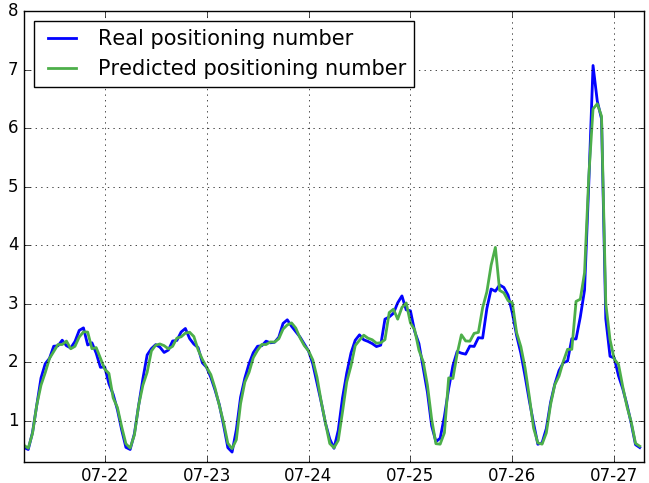} \\
(d) Forbidden City in 2015/05/01 & (e) Beijing National Stadium in 2015/10/07&
(f) Beijing Workers' Stadium in 2015/07/26\\
\end{tabular}
\caption{The predicted positioning number and real positioning number on the Baidu map corresponding to the aforementioned events.}
\label{fig:prediction}
\normalsize
\end{figure*}

For aforementioned 6 different POIs (Shanghai Bund, Shanghai Stadium, Shenzhen Bay Sports Center, Forbidden City, Beijing National Stadium, Beijing Workers' Stadium),
we train 6 distinctive positioning number prediction models for them based on the data in 60 days before the event in each POI.
As we can see from Figure \ref{fig:prediction}, the predicted positioning number of our model is very near to the real number.
That is to say, the model provides a reliable one hour ahead prediction in the 6 POIs.
We believe the positioning number prediction model can be extended to other POIs.

\begin{figure}[!htb]
\centering{}
\includegraphics[width=0.48 \textwidth]{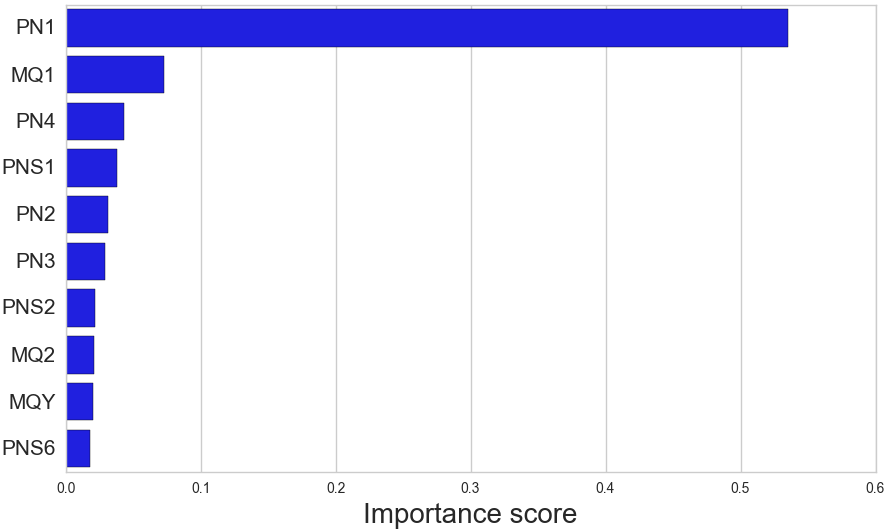}
\caption{The feature importance rankings in prediction model.} \label{fig:feature_importance}
\end{figure}

Furthermore, we evaluated the importance of all features in prediction model. The top 10 important features is illustrated in Figure \ref{fig:feature_importance} (feature description is in Table 1).
We calculate the Gini importance score for each feature: the higher the score is, the more important the feature is \cite{breiman2001random}.
As we can see from Figure \ref{fig:feature_importance}, ``PN1 (positioning number 1 hour ago)'' and ``MQ1 (map query number 1 hour ago)'' are two most important features in our model, which is consistent with our intuition.
It is worth noting that ``MQY (map query number in 20:00 - 24:00  yesterday)'' is also an importance feature as many people prefer to planing their routine in the night before.

\begin{figure}[!htb]
\centering{}
\includegraphics[width=0.46 \textwidth]{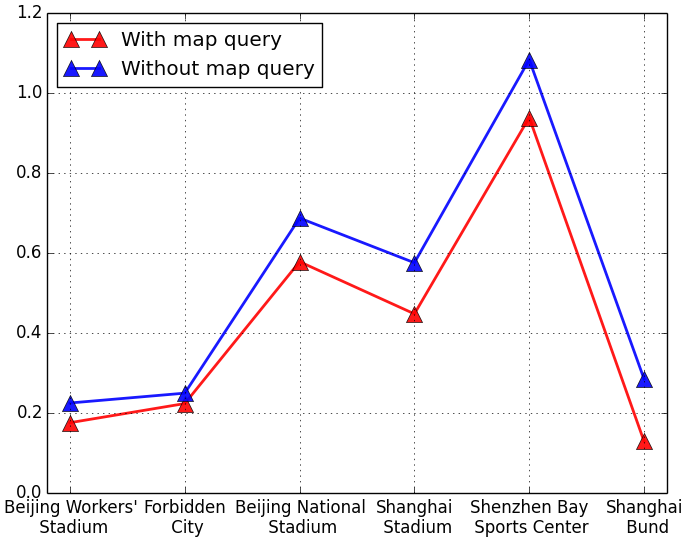}
\caption{The Mean Absolute Error (MAE) of the prediction model. Red line is the result of the model with features in Table \ref{tab:feature}, and blue line is the one with features in Table \ref{tab:feature} but without map query related features, and y denotes the MAE score} \label{fig:mae}
\end{figure}

From the importance ranking in Figure \ref{fig:feature_importance}, we can see features from map query data are a vital part for positioning number prediction.
In order to test the essentiality of map query data for the prediction, we further design a comparative  experiment where two prediction models are evaluated: one with features extracted from map query data and
the other without.
Then, we can directly capture the importance of map query data by quantity measure the prediction error of two models.
We use Mean Absolute Error (MAE) as evaluation metrics to measure how close predictions are to the real positioning numbers.
As the experiment result shown in Figure \ref{fig:mae}, the performance of model with the map query features is superior than the model without the map query features.
To sum up, map query features play a vital role in accurate positioning number prediction.